\documentclass[ALICE,manyauthors]{cernphprep}
\usepackage[comma,square,numbers,sort&compress]{natbib}
\usepackage{hyperref}
\usepackage{lineno}
\usepackage{xspace}
\usepackage{epstopdf}
\usepackage[T1]{fontenc}
\begin{document}
%%%%%%%%%%%%%%%%%%%%%%%%%%%%%%%%%%%%%%%%%%%%%%%%%%
% These are some new commands that may be useful 
% for paper writing in general. If other newcommands
% are needed for your specific paper, please feel 
% free to add here. 
%
% The currently available commands are organized in: 
% 1) Systems
% 2) Quantities
% 3) Energies and units
% 4) Detectors
% 5) particle species 
%%%%%%%%%%%%%%%%%%%%%%%%%%%%%%%%%%%%%%%%%%%%%%%%%%

% 1) SYSTEMS 
\newcommand{\pp}           {pp\xspace}
\newcommand{\ppbar}        {\mbox{$\mathrm {p\overline{p}}$}\xspace}
\newcommand{\XeXe}         {\mbox{Xe--Xe}\xspace}
\newcommand{\PbPb}         {\mbox{Pb--Pb}\xspace}
\newcommand{\pA}           {\mbox{pA}\xspace}
\newcommand{\pPb}          {\mbox{p--Pb}\xspace}
\newcommand{\AuAu}         {\mbox{Au--Au}\xspace}
\newcommand{\dAu}          {\mbox{d--Au}\xspace}

% 2) QUANTITIES 
\newcommand{\s}            {\ensuremath{\sqrt{s}}\xspace}
\newcommand{\snn}          {\ensuremath{\sqrt{s_{\mathrm{NN}}}}\xspace}
\newcommand{\pt}           {\ensuremath{p_{\rm T}}\xspace}
\newcommand{\meanpt}       {$\langle p_{\mathrm{T}}\rangle$\xspace}
\newcommand{\ycms}         {\ensuremath{y_{\rm CMS}}\xspace}
\newcommand{\ylab}         {\ensuremath{y_{\rm lab}}\xspace}
\newcommand{\etarange}[1]  {\mbox{$\left | \eta \right |~<~#1$}}
\newcommand{\yrange}[1]    {\mbox{$\left | y \right |~<~#1$}}
\newcommand{\dndy}         {\ensuremath{\mathrm{d}N_\mathrm{ch}/\mathrm{d}y}\xspace}
\newcommand{\dndeta}       {\ensuremath{\mathrm{d}N_\mathrm{ch}/\mathrm{d}\eta}\xspace}
\newcommand{\avdndeta}     {\ensuremath{\langle\dndeta\rangle}\xspace}
\newcommand{\dNdy}         {\ensuremath{\mathrm{d}N_\mathrm{ch}/\mathrm{d}y}\xspace}
\newcommand{\Npart}        {\ensuremath{N_\mathrm{part}}\xspace}
\newcommand{\Ncoll}        {\ensuremath{N_\mathrm{coll}}\xspace}
\newcommand{\dEdx}         {\ensuremath{\textrm{d}E/\textrm{d}x}\xspace}
\newcommand{\RpPb}         {\ensuremath{R_{\rm pPb}}\xspace}

% 3) ENERGIES, UNITS
\newcommand{\nineH}        {$\sqrt{s}~=~0.9$~Te\kern-.1emV\xspace}
\newcommand{\seven}        {$\sqrt{s}~=~7$~Te\kern-.1emV\xspace}
\newcommand{\twoH}         {$\sqrt{s}~=~0.2$~Te\kern-.1emV\xspace}
\newcommand{\twosevensix}  {$\sqrt{s}~=~2.76$~Te\kern-.1emV\xspace}
\newcommand{\five}         {$\sqrt{s}~=~5.02$~Te\kern-.1emV\xspace}
\newcommand{\twosevensixnn}{$\sqrt{s_{\mathrm{NN}}}~=~2.76$~Te\kern-.1emV\xspace}
\newcommand{\fivenn}       {$\sqrt{s_{\mathrm{NN}}}~=~5.02$~Te\kern-.1emV\xspace}
\newcommand{\LT}           {L{\'e}vy-Tsallis\xspace}
\newcommand{\GeVc}         {Ge\kern-.1emV/$c$\xspace}
\newcommand{\MeVc}         {Me\kern-.1emV/$c$\xspace}
\newcommand{\TeV}          {Te\kern-.1emV\xspace}
\newcommand{\GeV}          {Ge\kern-.1emV\xspace}
\newcommand{\MeV}          {Me\kern-.1emV\xspace}
\newcommand{\GeVmass}      {Ge\kern-.2emV/$c^2$\xspace}
\newcommand{\MeVmass}      {Me\kern-.2emV/$c^2$\xspace}
\newcommand{\lumi}         {\ensuremath{\mathcal{L}}\xspace}

% 4) DETECTORS 
\newcommand{\ITS}          {\rm{ITS}\xspace}
\newcommand{\TOF}          {\rm{TOF}\xspace}
\newcommand{\ZDC}          {\rm{ZDC}\xspace}
\newcommand{\ZDCs}         {\rm{ZDCs}\xspace}
\newcommand{\ZNA}          {\rm{ZNA}\xspace}
\newcommand{\ZNC}          {\rm{ZNC}\xspace}
\newcommand{\SPD}          {\rm{SPD}\xspace}
\newcommand{\SDD}          {\rm{SDD}\xspace}
\newcommand{\SSD}          {\rm{SSD}\xspace}
\newcommand{\TPC}          {\rm{TPC}\xspace}
\newcommand{\TRD}          {\rm{TRD}\xspace}
\newcommand{\VZERO}        {\rm{V0}\xspace}
\newcommand{\VZEROA}       {\rm{V0A}\xspace}
\newcommand{\VZEROC}       {\rm{V0C}\xspace}
\newcommand{\Vdecay} 	   {\ensuremath{V^{0}}\xspace}

% 4) PARTICLE SPECIES 
\newcommand{\ee}           {\ensuremath{e^{+}e^{-}}} 
\newcommand{\pip}          {\ensuremath{\pi^{+}}\xspace}
\newcommand{\pim}          {\ensuremath{\pi^{-}}\xspace}
\newcommand{\kap}          {\ensuremath{\rm{K}^{+}}\xspace}
\newcommand{\kam}          {\ensuremath{\rm{K}^{-}}\xspace}
\newcommand{\pbar}         {\ensuremath{\rm\overline{p}}\xspace}
\newcommand{\kzero}        {\ensuremath{{\rm K}^{0}_{\rm{S}}}\xspace}
\newcommand{\lmb}          {\ensuremath{\Lambda}\xspace}
\newcommand{\almb}         {\ensuremath{\overline{\Lambda}}\xspace}
\newcommand{\Om}           {\ensuremath{\Omega^-}\xspace}
\newcommand{\Mo}           {\ensuremath{\overline{\Omega}^+}\xspace}
\newcommand{\X}            {\ensuremath{\Xi^-}\xspace}
\newcommand{\Ix}           {\ensuremath{\overline{\Xi}^+}\xspace}
\newcommand{\Xis}          {\ensuremath{\Xi^{\pm}}\xspace}
\newcommand{\Oms}          {\ensuremath{\Omega^{\pm}}\xspace}
\newcommand{\degree}       {\ensuremath{^{\rm o}}\xspace}
\newcommand{\jpsi}         {\rm J/$\psi$}

%%%%%%%%%%%%%%%  Title page %%%%%%%%%%%%%%%%%%%%%%%%
\begin{titlepage}
% the dates below correspond to CERN approval
% please don't touch: EB chairs will take care
\PHyear{2019}       % required, will be obtained from CERN
\PHnumber{144}      % required, will be obtained from CERN
\PHdate{4 July}  % required, will be obtained from CERN
%%%%%%%%%%%%%%%%%%%%%%%%%%%%%%%%%%%%%%%%%%%%%%%%%%%%

%%% Put your own title + short title here:
\title{Measurement of $\Upsilon(1{\rm S})$ elliptic flow at forward rapidity in Pb--Pb collisions at
\fivenn}
%$\mathbf{\sqrt{{\textit s}_{\rm NN}}}$ = 5.02 TeV}
\ShortTitle{$\Upsilon$ $v_{2}$ in Pb--Pb collisions at \fivenn}   % appears on left page headers

%%% Do not change the next lines
\Collaboration{ALICE Collaboration\thanks{See Appendix~\ref{app:collab} for the list of collaboration members}}
\ShortAuthor{ALICE Collaboration} % appears on right page headers, do not change

\begin{abstract}
The first measurement of the $\Upsilon(1{\rm S})$ elliptic flow coefficient ($v_2$) is performed at forward 
rapidity (2.5~$<$~{\it y}~$<$~4) in Pb--Pb collisions at $\sqrt{s_{\rm NN}} = 5.02$~TeV with 
the ALICE detector at the LHC. The results are obtained with the scalar product method and are 
reported as a function of transverse momentum (\pt) up to 15 GeV/$c$ in the 5--60\% centrality interval. The measured $\Upsilon(1{\rm S})$ $v_2$ is consistent with zero 
and with the small positive values predicted by transport models within uncertainties.
The $v_2$ coefficient in 2~$<$~$p_{\rm T}$~$<$~15 GeV/$c$ is lower than that of inclusive \jpsi~mesons in the same \pt{} interval by 2.6 standard deviations. These results, combined with earlier suppression measurements, are in agreement with a scenario in which the $\Upsilon$(1S) production in Pb--Pb collisions at LHC energies is dominated by dissociation limited to the early stage of the collision whereas in the J/$\psi$ case there is substantial experimental evidence of an additional regeneration component.

\end{abstract}
\end{titlepage}

\setcounter{page}{2} %please do not remove this line

At the extreme energy densities and temperatures produced in ultra-relativistic collisions of heavy nuclei, hadronic matter undergoes a transition into a state of deconfined quarks and gluons, known as Quark--Gluon Plasma (QGP). The created QGP medium is characterized as a strongly coupled system, which behaves as an almost perfect fluid in the sense that its shear viscosity to entropy density ratio approaches the smallest possible values~\cite{Kovtun:2004de,Luzum:2008cw,Braun-Munzinger:2015hba}.
Spatial initial state anisotropy of the overlap region of the two colliding nuclei is transformed by the fluid pressure gradients into a momentum anisotropy of the produced final-state particles. This effect is known as hydrodynamic anisotropic flow~\cite{Ollitrault:1992bk} and is usually quantified in terms of the harmonic coefficients of
the Fourier decomposition of the azimuthal particle distribution~\cite{Voloshin:1994mz}. The dominant coefficient in non-central collisions is the second harmonic, denoted by $v_2$ and known as elliptic flow, since this coefficient directly arises from the almond-shaped interaction region between the colliding nuclei. It is approximately proportional to the eccentricity $\epsilon_{2}$ of the initial collision geometry~\cite{Gardim:2011xv}. The proportionality coefficient reflects the response of the QGP medium to the initial anisotropy and depends on the particle type, mass and kinematics~\cite{Qiu:2011iv}.

Charm and beauty quarks are important probes of the QGP. They are created predominantly in hard-scattering processes at the early collision stage and therefore experience the entire evolution of the QGP. The observed significant D-meson $v_2$ in nucleus--nucleus collisions suggests that the charm quarks participate in the collective anisotropic flow of the QGP fluid~\cite{Acharya:2017qps,Sirunyan:2017plt,Acharya:2018bxo}. Nevertheless, since the light-flavor quarks also contribute to the D-meson flow, detailed comparisons with theoretical models are necessary to draw firm conclusions about the charm-quark flow. Quarkonia, which are bound states of heavy-flavor quark-antiquark pairs, offer a complementary way to study the interaction of the heavy-flavor quarks with the medium and thus to independently shed light on the properties of the QGP~\cite{Andronic:2015wma}. In a simplified picture, quarkonium production is suppressed by color screening inside the QGP medium created in nucleus--nucleus collisions~\cite{Matsui:1986dk}. The level of suppression depends on the heavy-quark interaction and the temperature of the surrounding medium~\cite{Digal:2001ue,Mocsy:2013syh}. The azimuthal asymmetry of the overlap region of the two colliding nuclei and the dependence of the suppression on the path length traversed by the quark-antiquark pair inside the medium lead to positive $v_2$ values increasing as a function of transverse momentum ($p_{\rm T}$). At LHC energies, there is evidence for a competing effect which enhances the production of charmonia (bound states of charm quark-antiquark pairs)~\cite{Abelev:2012rv,Abelev:2013ila,Adam:2016rdg}. This effect originates from regeneration of charmonia via recombination of (partially) thermalized charm quarks either during the QGP evolution~\cite{He:2014cla,Du:2015wha} or at the QGP phase boundary~\cite{BraunMunzinger:2000px,Andronic:2018vqh}. It becomes significant at LHC energies due to the large charm-quark production cross section, which implies that a sufficiently high number of charm quarks traveling inside the QGP are available for recombination. Within the regeneration scenario, the elliptic flow of charmonia is directly inherited from the velocity field of the individual charm quarks within the medium and results in a positive $v_2$ coefficient, mainly at low $p_{\rm T}$. Measurements of significant ${\rm J}/\psi$-meson $v_2$ coefficient in Pb--Pb collisions at LHC energies clearly speaks in favor of charm-quark flow and the regeneration scenario~\cite{ALICE:2013xna,Acharya:2017tgv,Acharya:2018pjd,Aaboud:2018ttm}. Despite this, the phenomenological models which incorporate transport of heavy-flavor quark-antiquark pairs inside the QGP are not yet able to provide a fully satisfactory description of the $p_{\rm T}$ dependence of the measured J/$\psi$ elliptic flow~\cite{Zhou:2014kka,Du:2015wha}. Moreover, recent results in high-multiplicity p--Pb collisions also indicate a significant J/$\psi$ $v_2$~\cite{Acharya:2017tfn,Sirunyan:2018kiz}, which is unexpected within the present transport models due to the small collision-system size and low number of available charm quarks~\cite{Du:2018wsj}. Recent calculations within the Color-Glass Condensate framework attribute this significant $v_2$ to initial-state effects~\cite{Zhang:2019dth}.

Bottomonia, bound states of bottom quark-antiquark pairs, are also expected to be suppressed inside the QGP by the color-screening effect~\cite{Digal:2001ue,Brambilla:2010cs,Andronic:2015wma}. 
Indeed, measurements in Pb--Pb collisions at the LHC demonstrate a significant suppression of inclusive $\Upsilon$(1S) production~\cite{Chatrchyan:2012lxa,Abelev:2014nua,Khachatryan:2016xxp,Acharya:2018mni}.
In recent calculations the $v_2$ coefficient of inclusive $\Upsilon$(1S) is predicted to be significantly smaller when compared to that of inclusive J/$\psi$~\cite{Du:2017qkv}. The reason is that the $\Upsilon$(1S) dissociation happens at higher temperatures due to its greater binding energy. The dissociation is therefore limited to the earlier stage of the collision, when the path-length differences are less influential. In addition, the recombination of (partially) thermalized bottom quarks gives a negligible contribution to the $v_2$ coefficient due to the small number of available bottom quarks~\cite{Du:2017qkv}. As a result, the predicted values of $\Upsilon$(1S) $v_2$ coefficient are small in contrast to the charmonium case. It is worth noting that even though the $v_2$ coefficient of the excited bottomonium state $\Upsilon$(2S) is currently beyond experimental reach, it is expected to be significantly higher than that of $\Upsilon$(1S). Due to its lower binding energy and other bound-state characteristic differences, the suppression and regeneration occur up to a later stage of the collision. Hence, the path-length dependent suppression induces a larger $v_2$, the fraction of regenerated $\Upsilon$(2S) is higher and the inherited $v_2$ is larger~\cite{Du:2017qkv}.
Consequently, the measurement of the bottomonium elliptic flow is a crucial ingredient in the study of heavy-flavor interactions with the QGP, not only to complement the corresponding charmonium measurements, but also in the search for any sizable $v_2$ beyond the theoretical expectations.

In this Letter, we present the first measurement of $\Upsilon$(1S) elliptic flow in Pb--Pb collisions at $\sqrt{s_{\rm NN}}~=~5.02$ TeV at forward rapidity (2.5 $<$ $y$ $<$ 4). The $\Upsilon$ mesons are reconstructed via their $\mu^+\mu^-$ decay channel. The results are obtained in the momentum interval 0 $<$ $p_{\rm T}$ $<$ 15 GeV/$c$ and the 5--60\% collision centrality interval.

General information on the ALICE apparatus and its performance can be found 
in Refs.~\cite{Aamodt:2008zz, Abelev:2014ffa}. The muon spectrometer, which covers the pseudorapidity
range $-4 < \eta < -2.5$\footnote{In the ALICE reference frame, the muon spectrometer covers a negative $\eta$ range and consequently a negative $y$ range. The results were chosen to be presented with a positive $y$ notation, due to the symmetry of the collision system.}, is used to reconstruct muon tracks. It consists of a front absorber followed 
by five tracking stations with the third station placed inside a dipole magnet. Two trigger stations 
located downstream of an iron wall complete the spectrometer. The Silicon Pixel Detector 
(SPD)~\cite{Dellacasa:1999kf,Aamodt:2010aa} consists of two cylindrical layers covering the full azimuthal angle and $|\eta|$ $<$ 2.0 and $|\eta|$ $<$ 1.4, respectively. The SPD 
is employed to determine the position of the primary vertex and to reconstruct tracklets, track segments formed by the clusters in the two SPD layers and the primary vertex~\cite{Adam:2015gka}.
Two arrays of 32 scintillator counters each~\cite{Abbas:2013taa}, 
covering 2.8 $<$ $\eta$ $<$ 5.1 (V0A) and $-3.7 < \eta < -1.7$ (V0C), are used for triggering, 
the event selection and the determination of the collision centrality and the event flow vector. In addition, two neutron Zero Degree Calorimeters~\cite{Dellacasa:1999ke}, installed 112.5 m from the 
interaction point along the beam line on each side, are employed for the event selection. 

The data samples recorded by ALICE during the 2015 and 2018 LHC Pb--Pb runs at $\snn= 5.02$~TeV 
are used for this analysis. The trigger conditions and the event selection criteria are described 
in Ref.~\cite{Acharya:2018pjd}.
The primary vertex position is required to be within $\pm$14~cm 
from the nominal interaction point along the beam direction. The data are split in intervals of collision 
centrality, which is obtained based on the total signal in the V0A and V0C 
detectors~\cite{Adam:2015ptt}. The integrated luminosity of the analyzed data sample is about 750 $\mu$b$^{-1}$.

The muon selection is identical to that used in Refs.~\cite{Acharya:2017tfn,Acharya:2018pjd}. The dimuons are reconstructed 
in the acceptance of the muon spectrometer (2.5~$<$~{\it y}~$<$~4.0) and are required to have a transverse momentum between 0 and 15 GeV/$c$.
The alignment of the muon spectrometer is performed based on the MILLEPEDE package~\cite{Blobel:2002ax} and using Pb--Pb data taken with the nominal dipole magnetic field~\cite{Abelev:2014ffa}. The presence of the magnetic field limits the precision of the alignment procedure in the track bending direction. Indeed, a study of the reconstructed $\Upsilon$ mass as a function of the momentum of muon tracks ($p_{\mu}$) reveals a residual misalignment leading to a systematic shift in the measured muon track momentum $\Delta(1/p_{\mu})\approx\pm2.5\times10^{-4}$~(GeV/$c$)$^{-1}$, where the sign of the shift depends on the muon charge and the magnetic field polarity. A correction of this misalignment effect is obtained via a high-statistics sample of reconstructed J/$\psi\to\mu^+\mu^-$ decays and the spectra of high-momentum muon tracks. The correction is then applied to the reconstructed muon track momentum, resulting in up to 25\% improvement of the $\Upsilon$(1S) mass resolution for $p_{\rm T}$~$>$~6 GeV/$c$.

\begin{figure}[tb]
    \begin{center}
    \includegraphics[width = 0.48\textwidth]{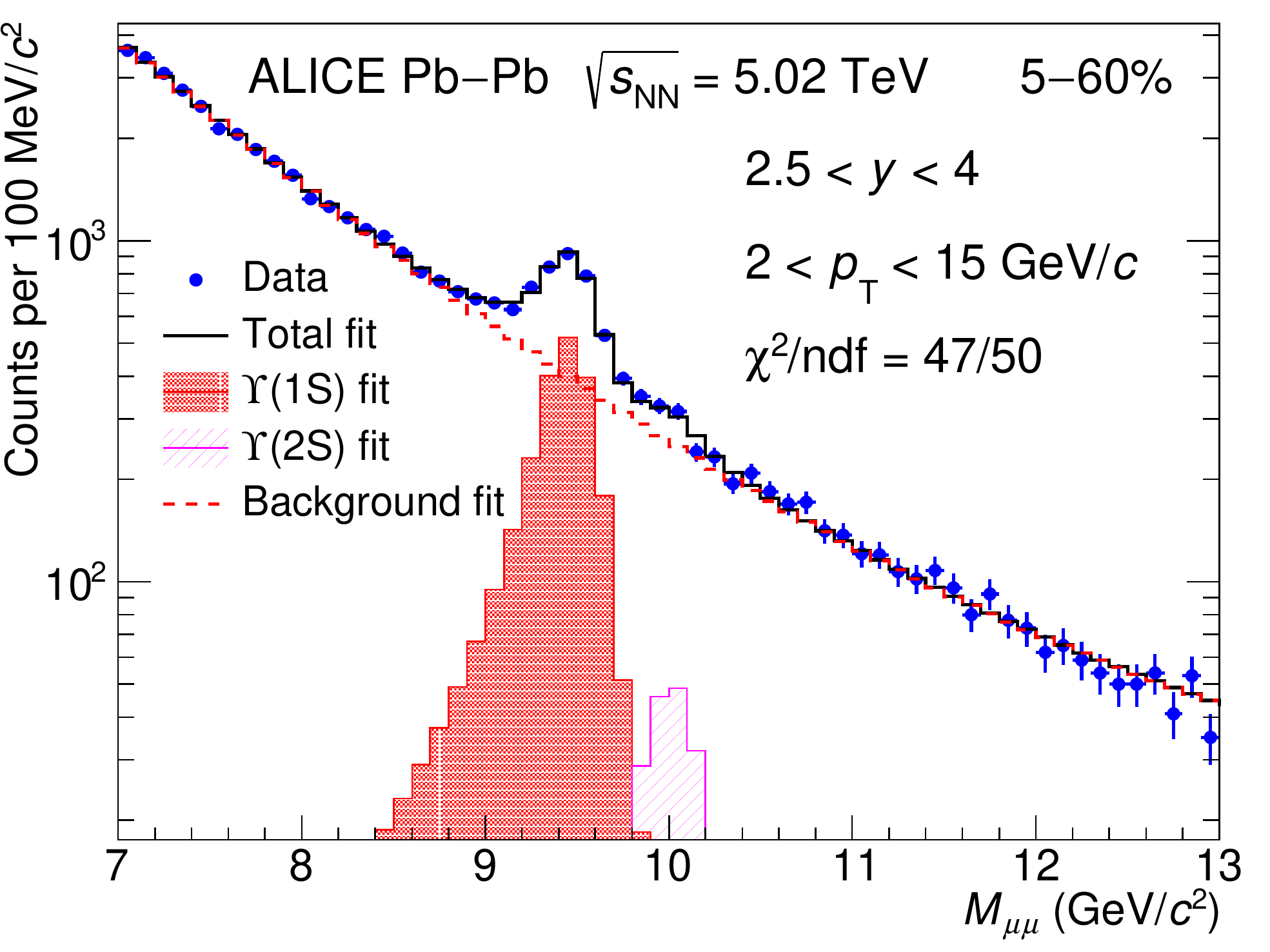}\hfill\includegraphics[width = 0.48\textwidth]{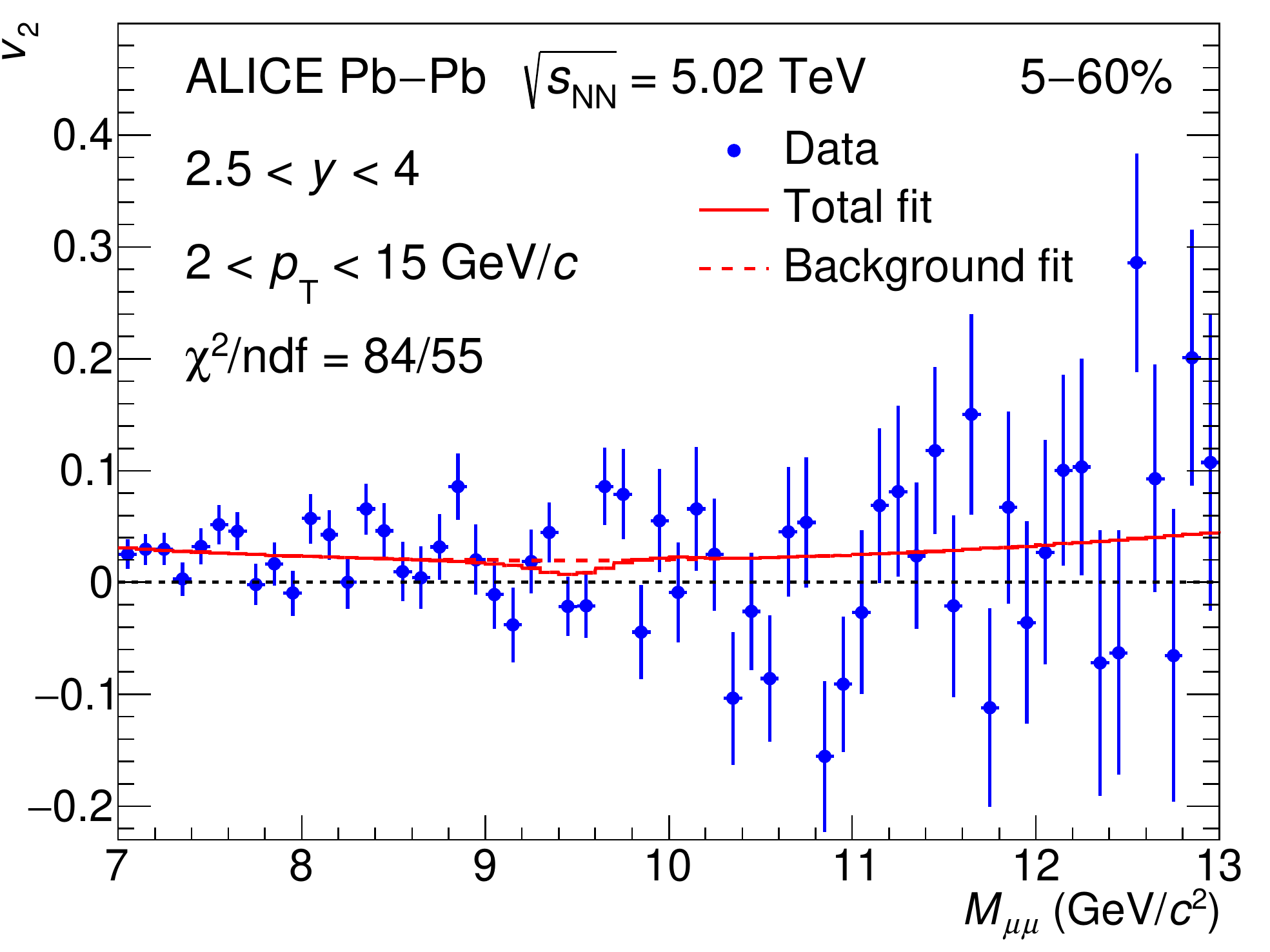}
    \end{center}
    \caption{(Color online) Left: The $M_{\mu\mu}$ distribution in the 5--60\% centrality interval and 2 $< \pt <$ 15 GeV/$c$ fitted with a combination of an
    extended Crystal Ball function for the signal and a Variable-Width Gaussian function for the background. Right: The $v_2(M_{\mu\mu})$ distribution
    in the same centrality and $p_{\rm T}$ intervals fitted with the function from Eq.~(\ref{eq:v2_sb}).}
    \label{fig:fits}
\end{figure}

The dimuon invariant mass ($M_{\mu\mu}$) distribution is
fitted with a combination of an extended Crystal Ball (CB2) function for the $\Upsilon$(1S) signal and a Variable-Width Gaussian (VWG) function with a quadratic dependence of the width on $M_{\mu\mu}$ for the background~\cite{ALICE-PUBLIC-2015-006}.
A binned maximum-likelihood fit is employed.
The $\Upsilon$(1S) peak position and width are left free, while the CB2 tail parameters are fixed to the values extracted from Monte Carlo simulations~\cite{Acharya:2018mni}. The $\Upsilon$(2S) and $\Upsilon$(3S) signals
are included in the fit. Their peak positions and widths are fixed to those of the $\Upsilon$(1S) scaled by the ratio of their nominal masses to the nominal mass of the $\Upsilon$(1S). An example of the $M_{\mu\mu}$ fit is shown in the left panel of Fig.~\ref{fig:fits}. It is worth noting that no statistically significant $\Upsilon$(3S) is observed in any of the studied centrality and \pt\ intervals, and thus it is not considered in the further analysis.

The dimuon $v_2$ is measured using the scalar product method~\cite{Adler:2002pu, Voloshin:2008dg}, correlating the reconstructed dimuons with the second-order harmonic event flow vector ${\bf Q}_2^{\rm SPD}$~\cite{Voloshin:1994mz, Barrette:1994xr} calculated from the azimuthal distribution of the reconstructed SPD tracklets
\begin{equation}
%    v_{2}\{{\rm SP}\} = \langle \langle {\bf u}_{2, k} {\bf Q}_2^{\rm SPD *} \rangle \rangle \Bigg/ \sqrt{ \frac{\langle {\bf Q}_2^{\rm SPD}  {\bf Q}_2^{\rm V0A *} \rangle \langle  {\bf Q}_2^{\rm SPD} {\bf Q}_2^{\rm V0C *} \rangle} { \langle  {\bf Q}_2^{\rm V0A} {\bf Q}_2^{\rm V0C *} \rangle } },
    v_{2}\{{\rm SP}\} = \Bigg\langle {\bf u}_{2} {\bf Q}_2^{\rm SPD *}\Bigg/ \sqrt{ \frac{\langle {\bf Q}_2^{\rm SPD}  {\bf Q}_2^{\rm V0A *} \rangle \langle  {\bf Q}_2^{\rm SPD} {\bf Q}_2^{\rm V0C *} \rangle} { \langle  {\bf Q}_2^{\rm V0A} {\bf Q}_2^{\rm V0C *} \rangle } } \Bigg\rangle_{\mu\mu},
    \label{eq:mth_sp}
\end{equation}
where ${\bf u}_{2} =\exp(i2\varphi)$ is the unit flow vector of the dimuon with azimuthal angle $\varphi$. The brackets $\langle \cdots \rangle_{\mu\mu}$ denote an average over all dimuons belonging to a given $\pt$, $M_{\mu\mu}$ and centrality interval. The ${\bf Q}_2^{\rm V0A}$ and ${\bf Q}_2^{\rm V0C}$ are the event flow vectors calculated from the azimuthal distribution of the energy deposition measured in the V0A and V0C detectors, respectively, and $^*$ is the complex conjugate. The brackets $\langle \cdots \rangle$ in the denominator denote an average over all events in a sufficiently narrow centrality class which encloses the event containing the dimuon. In order to account for a non-uniform detector response and efficiency, the components of all three event flow vectors are corrected using a recentering procedure~\cite{Selyuzhenkov:2007zi}. The gaps in pseudorapidity between the muon spectrometer and SPD ($|\Delta\eta|>1.0$) and between the SPD, V0A, and V0C remove auto-correlations and suppress short-range correlations unrelated to the azimuthal asymmetry in the initial geometry (``non-flow''), which largely come from jets and resonance decays. In the following, the $v_{2}\{{\rm SP}\} $ coefficient is denoted as $v_2$.

The $\Upsilon(1{\rm S})$ $v_2$ coefficient is obtained by a least squares fit of the superposition of the $\Upsilon$(1S) signal and the background to the dimuon flow coefficient as a function of the dimuon invariant mass~\cite{Borghini:2004ra}
\begin{equation}
%  v_{2}(M_{\mu\mu}) = \frac{N^{\Upsilon{\rm (1S)}}}{N^{\Upsilon{\rm (1S)}}+N^{\rm B}}v_{2}^{\Upsilon{\rm (1S)}} + \frac{N^{\rm B}}{N^{\Upsilon{\rm (1S)}}+N^{\rm B}}v_{2}^{\rm B}(M_{\mu\mu}),
  v_{2}(M_{\mu\mu}) = \alpha(M_{\mu\mu})v_{2}^{\Upsilon{\rm (1S)}} + [1-\alpha(M_{\mu\mu})]v_{2}^{\rm B}(M_{\mu\mu}),
  \label{eq:v2_sb}
\end{equation}
where $v_{2}^{\Upsilon{\rm (1S)}}$ is the flow coefficient of the signal, $v_{2}^{\rm B}$ is the $M_{\mu\mu}$-dependent flow coefficient of the background and $\alpha(M_{\mu\mu})$ is the signal fraction, obtained from the fit of the $M_{\mu\mu}$ distribution described above.
The background $v_2^{\rm B}$ is modeled as a second-order polynomial function of $M_{\mu\mu}$.
For consistency, and despite its low yield, the $\Upsilon$(2S) is included in the fit by restricting the value of its $v_2$ coefficient within the range between $-$0.5 and 0.5. In practice, this inclusion has a negligible impact on the $\Upsilon$(1S) fit results.
An example of $v_2$($M_{\mu\mu}$) fit is presented in the right panel of Fig.~\ref{fig:fits}.

The main systematic uncertainty of the measurement arises from the choice of the background fit function $v_2^{\rm B}$($M_{\mu\mu}$). In order to estimate this uncertainty, linear and constant functions are also used instead of the second-order polynomial. In addition, the signal CB2 tail parameters and background fit functions are varied~\cite{Acharya:2018mni}. The systematic uncertainty is then derived as the standard deviation with respect to the default choice of fitting functions. The absolute uncertainty increases from 0.004 to 0.016 with increasing collision centrality and decreasing $p_{\rm T}$, which is due to the decreasing signal-to-background ratio.
The dimuon trigger and reconstruction efficiency depends on the detector occupancy. This, coupled to the muon flow, could lead to a bias in the measured $v_2$. The corresponding systematic uncertainty is obtained by embedding simulated $\Upsilon$(1S) decays into real Pb--Pb events~\cite{Acharya:2018pjd}. It is found to be at most 0.0015 and is conservatively assumed to be the same in all transverse momentum and centrality intervals. The variations of the fit range and invariant-mass binning do not lead to deviations beyond the expected statistical fluctuations. The uncertainty related to the magnitude of the ${\bf Q}_2^{\rm SPD}$ flow vector is found to be negligible. Furthermore, the absence of any residual non-uniform detector acceptance and efficiency in the SPD flow vector determination after applying the recentering procedure is verified via the imaginary part of the scalar product (see Eq.~(\ref{eq:mth_sp}))~\cite{Selyuzhenkov:2007zi}.

\begin{figure}[tb]
    \begin{center}
    \includegraphics[width = 0.6\textwidth]{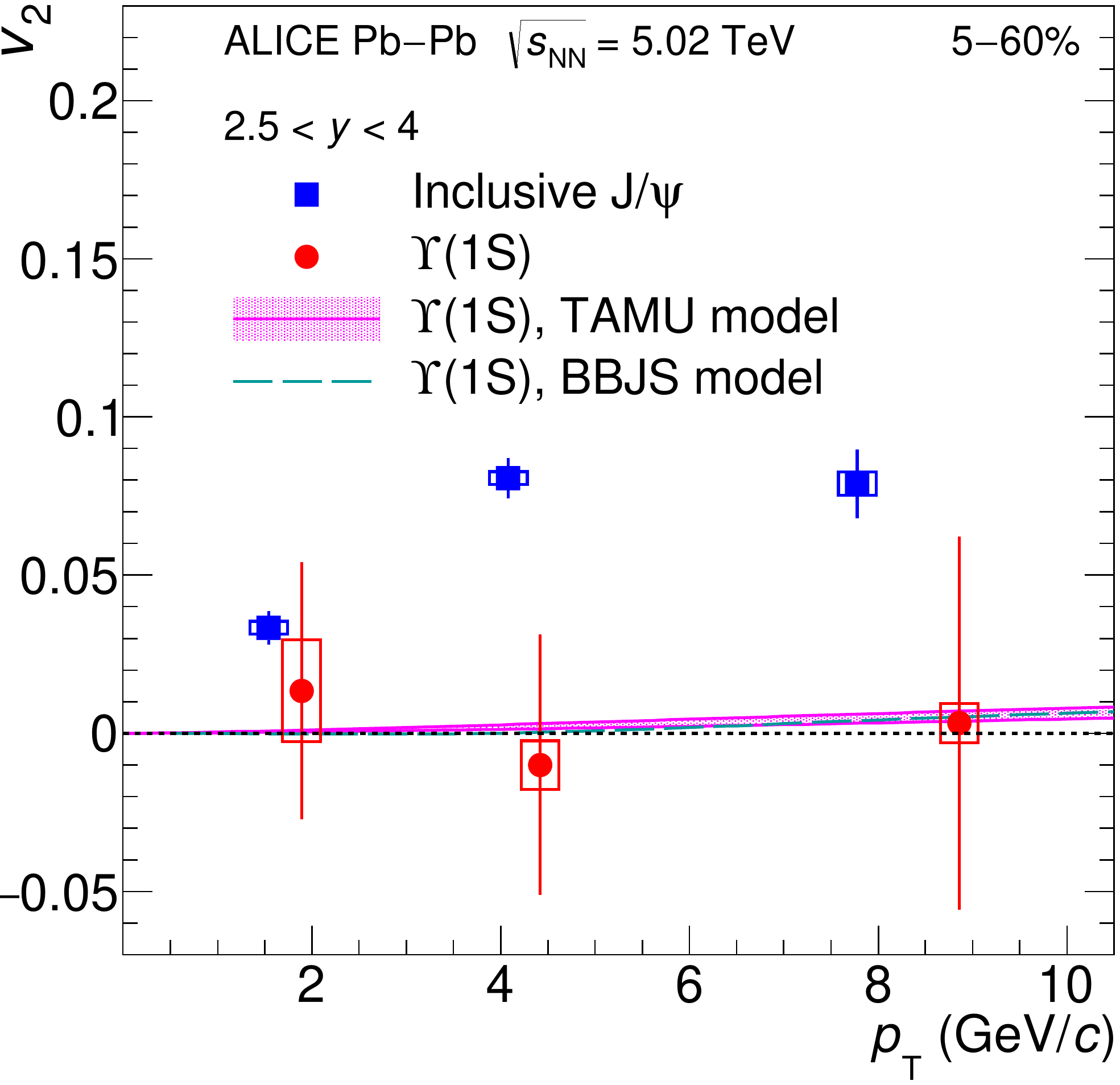}
    \end{center}
    \caption{(Color online) The $\Upsilon$(1S) $v_2$ coefficient as a function of \pt{} in the 5--60\% centrality interval compared to that of inclusive ${\rm J}/\psi$. The cyan dashed line represents the BBJS model calculations~\cite{Bhaduri:2018iwr}, while the magenta band denotes the TAMU model calculations~\cite{Du:2017qkv}. Error bars (open boxes) represent the statistical (systematic) uncertainties.}
    \label{fig:v2_pt}
\end{figure}

Figure~\ref{fig:v2_pt} shows the $\Upsilon$(1S) $v_2$ coefficient as a function of transverse momentum in the 5--60\% centrality interval. The central (0--5\%) and peripheral (60--100\%) collisions are not considered as the eccentricity of the initial collision geometry is small for the former and the signal yield is low in the latter. The $p_{\rm T}$ intervals are 0--3, 3--6, and 6--15 GeV/$c$ and the points are located at the average transverse momentum of the reconstructed $\Upsilon$(1S) uncorrected for detector acceptance and efficiency. The results are compatible with zero and with the small positive values predicted by the available theoretical models within uncertainties. The BBJS model calculations consider only the path-length dependent dissociation of initially-created bottomonia inside the QGP medium~\cite{Bhaduri:2018iwr}. The TAMU model incorporates in addition a regeneration component originating from the recombination of (partially) thermalized bottom quarks~\cite{Du:2017qkv}. Given that the regeneration component gives practically negligible contribution to the total $\Upsilon$(1S) $v_2$, the differences between the two models are marginal.
It is worth noting that although the quoted model predictions are for mid-rapidity, they remain valid also for the rapidity range of the measurement within the theoretical uncertainties. Indeed the fractions of regenerated and initially-produced $\Upsilon$(1S) are very close at mid- and forward rapidities~\cite{Du:2017qkv}. In addition, the QGP medium evolution is also similar between mid- and forward rapidities, given the weak rapidity dependence of the charged-particle multiplicity density~\cite{Adam:2016ddh}.
The presented $\Upsilon$(1S) $v_2$ result is coherent with the measured $\Upsilon$(1S) suppression in Pb--Pb collisions~\cite{Acharya:2018mni}, as the level of suppression is also fairly well reproduced by the BBJS model and the TAMU model including or excluding a regeneration component.
Therefore, the result is in agreement with a scenario in which the predominant mechanism affecting $\Upsilon$(1S) production in Pb--Pb collisions at the LHC energies is the dissociation limited to the early stage of the collision. It is interesting to note that the presented $\Upsilon$(1S) $v_2$ results are reminiscent of the corresponding charmonia measurements in Au--Au collisions at RHIC~\cite{Adamczyk:2012pw}, where so far non-observation of significant $v_2$ is commonly interpreted as a sign of a small regeneration component from recombination of thermalized charm quarks at lower RHIC energies.

The $\Upsilon$(1S) $v_2$ values in the three $p_{\rm T}$ intervals shown in Fig.~\ref{fig:v2_pt} are found to be lower, albeit with large uncertainties, compared to those of the inclusive ${\rm J}/\psi$ measured in the same centrality and $p_{\rm T}$ intervals using the data sample and analysis procedure described in Ref.~\cite{Acharya:2018pjd}.

\begin{figure}[tb]
    \begin{center}
    \includegraphics[width = 0.6\textwidth]{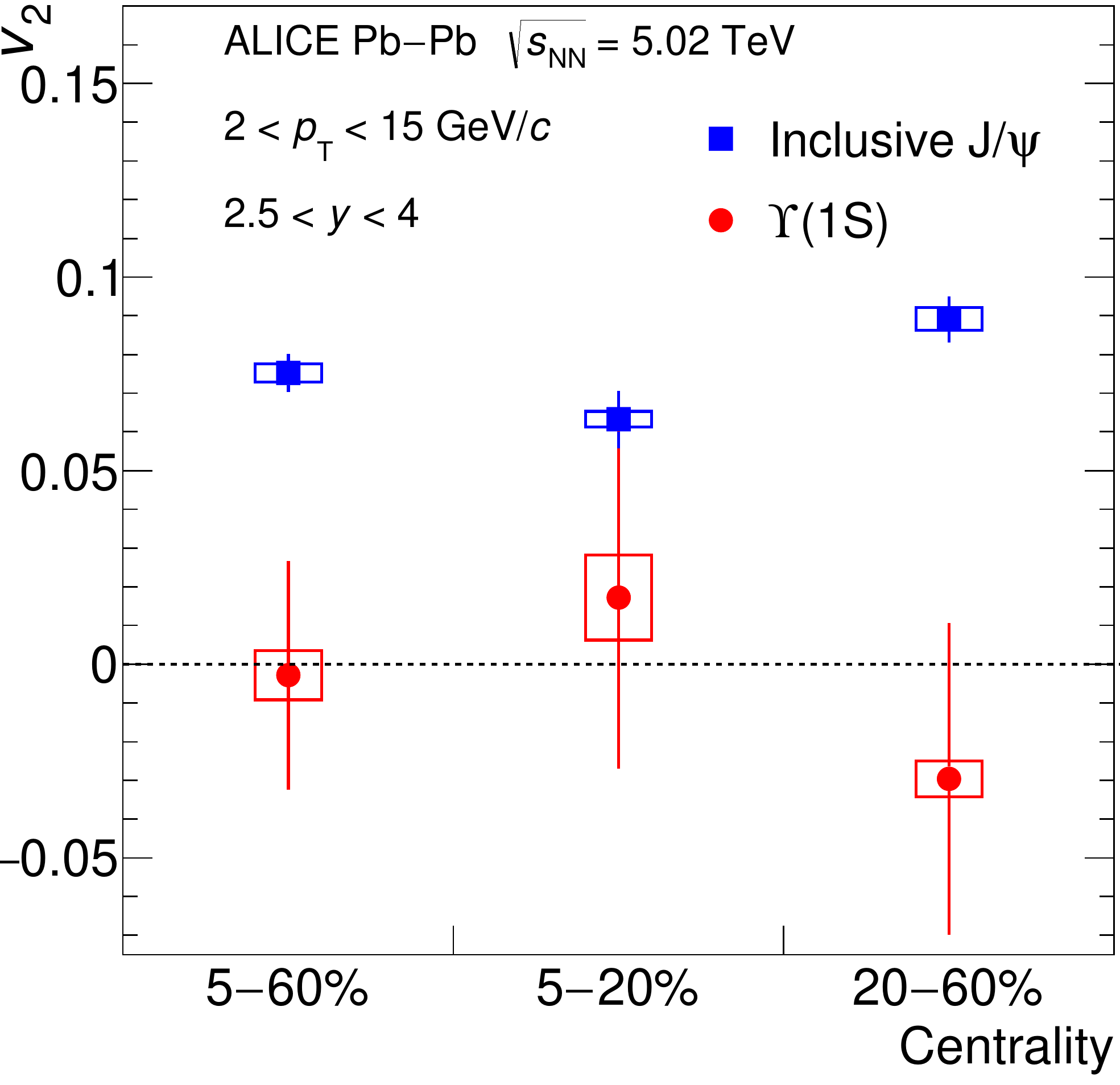}
    \end{center}
    \caption{(Color online) The $\Upsilon$(1S) $v_2$ coefficient integrated over the transverse momentum range 2 $< \pt <$ 15 GeV/$c$ in three centrality intervals compared to that of inclusive ${\rm J}/\psi$. Error bars (open boxes) represent the statistical (systematic) uncertainties.}
    \label{fig:v2_int_cen}
\end{figure}
Given that any $v_2$ originating either from recombination or from path-length dependent dissociation vanishes at zero \pt{}, the observed difference between $\Upsilon$(1S) and J/$\psi$ $v_2$ is quantified by performing the \pt-integrated measurement excluding the low \pt{} range. Figure~\ref{fig:v2_int_cen} presents the $\Upsilon$(1S) $v_2$ coefficient integrated over the transverse momentum range $2 < \pt < 15$~GeV/$c$ for three centrality intervals compared with that of the inclusive ${\rm J}/\psi$. The $\Upsilon$(1S) $v_2$ is found to be $-0.003 \pm 0.030(\rm{stat}) \pm 0.006 (\rm{syst})$ in the $2 < \pt < 15$~GeV/$c$ and 5--60\% centrality interval. This value is lower than the corresponding ${\rm J}/\psi$ $v_2$ by 2.6$\sigma$. This observation, coupled to the different measured centrality and $p_{\rm T}$ dependence of the $\Upsilon$(1S) and J/$\psi$ suppression in Pb--Pb collisions at the LHC~\cite{Acharya:2018mni,Adam:2016rdg}, can be interpreted within the models used for comparison as a sign that unlike $\Upsilon$(1S), J/$\psi$ production has a significant regeneration component. Nevertheless, no firm conclusions can be drawn, given that currently the transport models can not explain the significant J/$\psi$ $v_2$ for $p_{\rm T}$~$>$~4-5 GeV/$c$ observed in the data~\cite{Acharya:2017tgv}.

In summary, the first measurement of the $\Upsilon$(1S) $v_2$ coefficient in Pb--Pb collisions at $\sqrt{s_{\rm NN}} = 5.02$~TeV is presented. The measurement is performed in the 5--60\% centrality interval within $0 < \pt < 15$~GeV/$c$ range at forward rapidity. The $v_2$ coefficient is compatible with zero and with the model predictions within uncertainties. 
Excluding low $\pt$ ($0 < p_{\rm T} < 2$~GeV/$c$), $\Upsilon$(1S) $v_2$ is found to be 2.6$\sigma$ lower with respect to that of inclusive ${\rm J}/\psi$. The presented measurement opens the way for further studies of bottomonium flow using the future data samples from the LHC Runs 3 and 4 with an expected ten-fold increase in the number of the $\Upsilon$ candidates~\cite{ALICE-PUBLIC-2019-001,Citron:2018lsq}.

%%%%% acknowledgements - handled by EB chairs 
\newenvironment{acknowledgement}{\relax}{\relax}
\begin{acknowledgement}
\section*{Acknowledgements}
% add specific acknowledgements here 
% ...but please don't remove the line below: funding agencies
% will be acknowledged with a custom tex file handled by EB chairs after Collab Round 2
% Version: 2019-07-20

The ALICE Collaboration would like to thank all its engineers and technicians for their invaluable contributions to the construction of the experiment and the CERN accelerator teams for the outstanding performance of the LHC complex.
The ALICE Collaboration gratefully acknowledges the resources and support provided by all Grid centres and the Worldwide LHC Computing Grid (WLCG) collaboration.
The ALICE Collaboration acknowledges the following funding agencies for their support in building and running the ALICE detector:
A. I. Alikhanyan National Science Laboratory (Yerevan Physics Institute) Foundation (ANSL), State Committee of Science and World Federation of Scientists (WFS), Armenia;
Austrian Academy of Sciences, Austrian Science Fund (FWF): [M 2467-N36] and Nationalstiftung f\"{u}r Forschung, Technologie und Entwicklung, Austria;
Ministry of Communications and High Technologies, National Nuclear Research Center, Azerbaijan;
Conselho Nacional de Desenvolvimento Cient\'{\i}fico e Tecnol\'{o}gico (CNPq), Universidade Federal do Rio Grande do Sul (UFRGS), Financiadora de Estudos e Projetos (Finep) and Funda\c{c}\~{a}o de Amparo \`{a} Pesquisa do Estado de S\~{a}o Paulo (FAPESP), Brazil;
Ministry of Science \& Technology of China (MSTC), National Natural Science Foundation of China (NSFC) and Ministry of Education of China (MOEC) , China;
Croatian Science Foundation and Ministry of Science and Education, Croatia;
Centro de Aplicaciones Tecnol\'{o}gicas y Desarrollo Nuclear (CEADEN), Cubaenerg\'{\i}a, Cuba;
Ministry of Education, Youth and Sports of the Czech Republic, Czech Republic;
The Danish Council for Independent Research | Natural Sciences, the Carlsberg Foundation and Danish National Research Foundation (DNRF), Denmark;
Helsinki Institute of Physics (HIP), Finland;
Commissariat \`{a} l'Energie Atomique (CEA), Institut National de Physique Nucl\'{e}aire et de Physique des Particules (IN2P3) and Centre National de la Recherche Scientifique (CNRS) and R\'{e}gion des  Pays de la Loire, France;
Bundesministerium f\"{u}r Bildung und Forschung (BMBF) and GSI Helmholtzzentrum f\"{u}r Schwerionenforschung GmbH, Germany;
General Secretariat for Research and Technology, Ministry of Education, Research and Religions, Greece;
National Research, Development and Innovation Office, Hungary;
Department of Atomic Energy Government of India (DAE), Department of Science and Technology, Government of India (DST), University Grants Commission, Government of India (UGC) and Council of Scientific and Industrial Research (CSIR), India;
Indonesian Institute of Science, Indonesia;
Centro Fermi - Museo Storico della Fisica e Centro Studi e Ricerche Enrico Fermi and Istituto Nazionale di Fisica Nucleare (INFN), Italy;
Institute for Innovative Science and Technology , Nagasaki Institute of Applied Science (IIST), Japan Society for the Promotion of Science (JSPS) KAKENHI and Japanese Ministry of Education, Culture, Sports, Science and Technology (MEXT), Japan;
Consejo Nacional de Ciencia (CONACYT) y Tecnolog\'{i}a, through Fondo de Cooperaci\'{o}n Internacional en Ciencia y Tecnolog\'{i}a (FONCICYT) and Direcci\'{o}n General de Asuntos del Personal Academico (DGAPA), Mexico;
Nederlandse Organisatie voor Wetenschappelijk Onderzoek (NWO), Netherlands;
The Research Council of Norway, Norway;
Commission on Science and Technology for Sustainable Development in the South (COMSATS), Pakistan;
Pontificia Universidad Cat\'{o}lica del Per\'{u}, Peru;
Ministry of Science and Higher Education and National Science Centre, Poland;
Korea Institute of Science and Technology Information and National Research Foundation of Korea (NRF), Republic of Korea;
Ministry of Education and Scientific Research, Institute of Atomic Physics and Ministry of Research and Innovation and Institute of Atomic Physics, Romania;
Joint Institute for Nuclear Research (JINR), Ministry of Education and Science of the Russian Federation, National Research Centre Kurchatov Institute, Russian Science Foundation and Russian Foundation for Basic Research, Russia;
Ministry of Education, Science, Research and Sport of the Slovak Republic, Slovakia;
National Research Foundation of South Africa, South Africa;
Swedish Research Council (VR) and Knut \& Alice Wallenberg Foundation (KAW), Sweden;
European Organization for Nuclear Research, Switzerland;
National Science and Technology Development Agency (NSDTA), Suranaree University of Technology (SUT) and Office of the Higher Education Commission under NRU project of Thailand, Thailand;
Turkish Atomic Energy Agency (TAEK), Turkey;
National Academy of  Sciences of Ukraine, Ukraine;
Science and Technology Facilities Council (STFC), United Kingdom;
National Science Foundation of the United States of America (NSF) and United States Department of Energy, Office of Nuclear Physics (DOE NP), United States of America.
\end{acknowledgement}

%%%%%%%% Bibliography 
\bibliographystyle{utphys}   % Remember we use title in the biblio
\bibliography{bibliography}
%\input {bibliography.tex}  

%%%%%%%%%%%%%%%%%%%%%%%%%%%%%%%%
% Appendices: yours (if any) + authorlist
%%%%%%%%%%%%%%%%%%%%%%%%%%%%%%%%
\newpage
\appendix

%
%\input{} % put your appendices here (if any)
%

%%%%% Authorlist - please do not touch: handled by EB chairs 
\section{The ALICE Collaboration}
\label{app:collab}
% Collaboration: CERN-LHC-ALICE
% Generation Date is 2019-Jun-18

% How to use:
%%%%%%%%% appendix with author list
%\appendix
%\section{The ALICE Collaboration}
%\label{app:collab}
%\input{Alice_Authorslist_XXXX-Axx-XX.tex}
\begingroup
\small
\begin{flushleft}
S.~Acharya\Irefn{org141}\And 
D.~Adamov\'{a}\Irefn{org93}\And 
S.P.~Adhya\Irefn{org141}\And 
A.~Adler\Irefn{org73}\And 
J.~Adolfsson\Irefn{org79}\And 
M.M.~Aggarwal\Irefn{org98}\And 
G.~Aglieri Rinella\Irefn{org34}\And 
M.~Agnello\Irefn{org31}\And 
N.~Agrawal\Irefn{org10}\textsuperscript{,}\Irefn{org48}\textsuperscript{,}\Irefn{org53}\And 
Z.~Ahammed\Irefn{org141}
% ...
\And
\renewcommand\labelenumi{\textsuperscript{\theenumi}~}

\section*{Affiliation notes}
\renewcommand\theenumi{\roman{enumi}}
\begin{Authlist}
\item \Adef{org*}Deceased
\item \Adef{orgI}Dipartimento DET del Politecnico di Torino, Turin, Italy
\item \Adef{orgII}M.V. Lomonosov Moscow State University, D.V. Skobeltsyn Institute of Nuclear, Physics, Moscow, Russia
\item \Adef{orgIII}Department of Applied Physics, Aligarh Muslim University, Aligarh, India
\item \Adef{orgIV}Institute of Theoretical Physics, University of Wroclaw, Poland
\end{Authlist}

\section*{Collaboration Institutes}
\renewcommand\theenumi{\arabic{enumi}~}
\begin{Authlist}
\item \Idef{org1}A.I. Alikhanyan National Science Laboratory (Yerevan Physics Institute) Foundation, Yerevan, Armenia
\item \Idef{org2}Bogolyubov Institute for Theoretical Physics, National Academy of Sciences of Ukraine, Kiev, Ukraine
\item \Idef{org3}Bose Institute, Department of Physics  and Centre for Astroparticle Physics and Space Science (CAPSS), Kolkata, India
\item \Idef{org4}Budker Institute for Nuclear Physics, Novosibirsk, Russia
\item \Idef{org5}California Polytechnic State University, San Luis Obispo, California, United States
\item \Idef{org6}Central China Normal University, Wuhan, China
\item \Idef{org7}Centre de Calcul de l'IN2P3, Villeurbanne, Lyon, France
\item \Idef{org8}Centro de Aplicaciones Tecnol\'{o}gicas y Desarrollo Nuclear (CEADEN), Havana, Cuba
\item \Idef{org9}Centro de Investigaci\'{o}n y de Estudios Avanzados (CINVESTAV), Mexico City and M\'{e}rida, Mexico
\item \Idef{org10}Centro Fermi - Museo Storico della Fisica e Centro Studi e Ricerche ``Enrico Fermi', Rome, Italy
\item \Idef{org11}Chicago State University, Chicago, Illinois, United States
\item \Idef{org12}China Institute of Atomic Energy, Beijing, China
\item \Idef{org13}Chonbuk National University, Jeonju, Republic of Korea
\item \Idef{org14}Comenius University Bratislava, Faculty of Mathematics, Physics and Informatics, Bratislava, Slovakia
\item \Idef{org15}COMSATS University Islamabad, Islamabad, Pakistan
\item \Idef{org16}Creighton University, Omaha, Nebraska, United States
\item \Idef{org17}Department of Physics, Aligarh Muslim University, Aligarh, India
\item \Idef{org18}Department of Physics, Pusan National University, Pusan, Republic of Korea
\item \Idef{org19}Department of Physics, Sejong University, Seoul, Republic of Korea
\item \Idef{org20}Department of Physics, University of California, Berkeley, California, United States
\item \Idef{org21}Department of Physics, University of Oslo, Oslo, Norway
\item \Idef{org22}Department of Physics and Technology, University of Bergen, Bergen, Norway
\item \Idef{org23}Dipartimento di Fisica dell'Universit\`{a} 'La Sapienza' and Sezione INFN, Rome, Italy
\item \Idef{org24}Dipartimento di Fisica dell'Universit\`{a} and Sezione INFN, Cagliari, Italy
\item \Idef{org25}Dipartimento di Fisica dell'Universit\`{a} and Sezione INFN, Trieste, Italy
\item \Idef{org26}Dipartimento di Fisica dell'Universit\`{a} and Sezione INFN, Turin, Italy
\item \Idef{org27}Dipartimento di Fisica e Astronomia dell'Universit\`{a} and Sezione INFN, Bologna, Italy
\item \Idef{org28}Dipartimento di Fisica e Astronomia dell'Universit\`{a} and Sezione INFN, Catania, Italy
\item \Idef{org29}Dipartimento di Fisica e Astronomia dell'Universit\`{a} and Sezione INFN, Padova, Italy
\item \Idef{org30}Dipartimento di Fisica `E.R.~Caianiello' dell'Universit\`{a} and Gruppo Collegato INFN, Salerno, Italy
\item \Idef{org31}Dipartimento DISAT del Politecnico and Sezione INFN, Turin, Italy
\item \Idef{org32}Dipartimento di Scienze e Innovazione Tecnologica dell'Universit\`{a} del Piemonte Orientale and INFN Sezione di Torino, Alessandria, Italy
\item \Idef{org33}Dipartimento Interateneo di Fisica `M.~Merlin' and Sezione INFN, Bari, Italy
\item \Idef{org34}European Organization for Nuclear Research (CERN), Geneva, Switzerland
\item \Idef{org35}Faculty of Electrical Engineering, Mechanical Engineering and Naval Architecture, University of Split, Split, Croatia
\item \Idef{org36}Faculty of Engineering and Science, Western Norway University of Applied Sciences, Bergen, Norway
\item \Idef{org37}Faculty of Nuclear Sciences and Physical Engineering, Czech Technical University in Prague, Prague, Czech Republic
\item \Idef{org38}Faculty of Science, P.J.~\v{S}af\'{a}rik University, Ko\v{s}ice, Slovakia
\item \Idef{org39}Frankfurt Institute for Advanced Studies, Johann Wolfgang Goethe-Universit\"{a}t Frankfurt, Frankfurt, Germany
\item \Idef{org40}Gangneung-Wonju National University, Gangneung, Republic of Korea
\item \Idef{org41}Gauhati University, Department of Physics, Guwahati, India
\item \Idef{org42}Helmholtz-Institut f\"{u}r Strahlen- und Kernphysik, Rheinische Friedrich-Wilhelms-Universit\"{a}t Bonn, Bonn, Germany
\item \Idef{org43}Helsinki Institute of Physics (HIP), Helsinki, Finland
\item \Idef{org44}High Energy Physics Group,  Universidad Aut\'{o}noma de Puebla, Puebla, Mexico
\item \Idef{org45}Hiroshima University, Hiroshima, Japan
\item \Idef{org46}Hochschule Worms, Zentrum  f\"{u}r Technologietransfer und Telekommunikation (ZTT), Worms, Germany
\item \Idef{org47}Horia Hulubei National Institute of Physics and Nuclear Engineering, Bucharest, Romania
\item \Idef{org48}Indian Institute of Technology Bombay (IIT), Mumbai, India
\item \Idef{org49}Indian Institute of Technology Indore, Indore, India
\item \Idef{org50}Indonesian Institute of Sciences, Jakarta, Indonesia
\item \Idef{org51}INFN, Laboratori Nazionali di Frascati, Frascati, Italy
\item \Idef{org52}INFN, Sezione di Bari, Bari, Italy
\item \Idef{org53}INFN, Sezione di Bologna, Bologna, Italy
\item \Idef{org54}INFN, Sezione di Cagliari, Cagliari, Italy
\item \Idef{org55}INFN, Sezione di Catania, Catania, Italy
\item \Idef{org56}INFN, Sezione di Padova, Padova, Italy
\item \Idef{org57}INFN, Sezione di Roma, Rome, Italy
\item \Idef{org58}INFN, Sezione di Torino, Turin, Italy
\item \Idef{org59}INFN, Sezione di Trieste, Trieste, Italy
\item \Idef{org60}Inha University, Incheon, Republic of Korea
\item \Idef{org61}Institut de Physique Nucl\'{e}aire d'Orsay (IPNO), Institut National de Physique Nucl\'{e}aire et de Physique des Particules (IN2P3/CNRS), Universit\'{e} de Paris-Sud, Universit\'{e} Paris-Saclay, Orsay, France
\item \Idef{org62}Institute for Nuclear Research, Academy of Sciences, Moscow, Russia
\item \Idef{org63}Institute for Subatomic Physics, Utrecht University/Nikhef, Utrecht, Netherlands
\item \Idef{org64}Institute of Experimental Physics, Slovak Academy of Sciences, Ko\v{s}ice, Slovakia
\item \Idef{org65}Institute of Physics, Homi Bhabha National Institute, Bhubaneswar, India
\item \Idef{org66}Institute of Physics of the Czech Academy of Sciences, Prague, Czech Republic
\item \Idef{org67}Institute of Space Science (ISS), Bucharest, Romania
\item \Idef{org68}Institut f\"{u}r Kernphysik, Johann Wolfgang Goethe-Universit\"{a}t Frankfurt, Frankfurt, Germany
\item \Idef{org69}Instituto de Ciencias Nucleares, Universidad Nacional Aut\'{o}noma de M\'{e}xico, Mexico City, Mexico
\item \Idef{org70}Instituto de F\'{i}sica, Universidade Federal do Rio Grande do Sul (UFRGS), Porto Alegre, Brazil
\item \Idef{org71}Instituto de F\'{\i}sica, Universidad Nacional Aut\'{o}noma de M\'{e}xico, Mexico City, Mexico
\item \Idef{org72}iThemba LABS, National Research Foundation, Somerset West, South Africa
\item \Idef{org73}Johann-Wolfgang-Goethe Universit\"{a}t Frankfurt Institut f\"{u}r Informatik, Fachbereich Informatik und Mathematik, Frankfurt, Germany
\item \Idef{org74}Joint Institute for Nuclear Research (JINR), Dubna, Russia
\item \Idef{org75}Korea Institute of Science and Technology Information, Daejeon, Republic of Korea
\item \Idef{org76}KTO Karatay University, Konya, Turkey
\item \Idef{org77}Laboratoire de Physique Subatomique et de Cosmologie, Universit\'{e} Grenoble-Alpes, CNRS-IN2P3, Grenoble, France
\item \Idef{org78}Lawrence Berkeley National Laboratory, Berkeley, California, United States
\item \Idef{org79}Lund University Department of Physics, Division of Particle Physics, Lund, Sweden
\item \Idef{org80}Nagasaki Institute of Applied Science, Nagasaki, Japan
\item \Idef{org81}Nara Women{'}s University (NWU), Nara, Japan
\item \Idef{org82}National and Kapodistrian University of Athens, School of Science, Department of Physics , Athens, Greece
\item \Idef{org83}National Centre for Nuclear Research, Warsaw, Poland
\item \Idef{org84}National Institute of Science Education and Research, Homi Bhabha National Institute, Jatni, India
\item \Idef{org85}National Nuclear Research Center, Baku, Azerbaijan
\item \Idef{org86}National Research Centre Kurchatov Institute, Moscow, Russia
\item \Idef{org87}Niels Bohr Institute, University of Copenhagen, Copenhagen, Denmark
\item \Idef{org88}Nikhef, National institute for subatomic physics, Amsterdam, Netherlands
\item \Idef{org89}NRC Kurchatov Institute IHEP, Protvino, Russia
\item \Idef{org90}NRC «Kurchatov Institute»  - ITEP, Moscow, Russia
\item \Idef{org91}NRNU Moscow Engineering Physics Institute, Moscow, Russia
\item \Idef{org92}Nuclear Physics Group, STFC Daresbury Laboratory, Daresbury, United Kingdom
\item \Idef{org93}Nuclear Physics Institute of the Czech Academy of Sciences, \v{R}e\v{z} u Prahy, Czech Republic
\item \Idef{org94}Oak Ridge National Laboratory, Oak Ridge, Tennessee, United States
\item \Idef{org95}Ohio State University, Columbus, Ohio, United States
\item \Idef{org96}Petersburg Nuclear Physics Institute, Gatchina, Russia
\item \Idef{org97}Physics department, Faculty of science, University of Zagreb, Zagreb, Croatia
\item \Idef{org98}Physics Department, Panjab University, Chandigarh, India
\item \Idef{org99}Physics Department, University of Jammu, Jammu, India
\item \Idef{org100}Physics Department, University of Rajasthan, Jaipur, India
\item \Idef{org101}Physikalisches Institut, Eberhard-Karls-Universit\"{a}t T\"{u}bingen, T\"{u}bingen, Germany
\item \Idef{org102}Physikalisches Institut, Ruprecht-Karls-Universit\"{a}t Heidelberg, Heidelberg, Germany
\item \Idef{org103}Physik Department, Technische Universit\"{a}t M\"{u}nchen, Munich, Germany
\item \Idef{org104}Politecnico di Bari, Bari, Italy
\item \Idef{org105}Research Division and ExtreMe Matter Institute EMMI, GSI Helmholtzzentrum f\"ur Schwerionenforschung GmbH, Darmstadt, Germany
\item \Idef{org106}Rudjer Bo\v{s}kovi\'{c} Institute, Zagreb, Croatia
\item \Idef{org107}Russian Federal Nuclear Center (VNIIEF), Sarov, Russia
\item \Idef{org108}Saha Institute of Nuclear Physics, Homi Bhabha National Institute, Kolkata, India
\item \Idef{org109}School of Physics and Astronomy, University of Birmingham, Birmingham, United Kingdom
\item \Idef{org110}Secci\'{o}n F\'{\i}sica, Departamento de Ciencias, Pontificia Universidad Cat\'{o}lica del Per\'{u}, Lima, Peru
\item \Idef{org111}Shanghai Institute of Applied Physics, Shanghai, China
\item \Idef{org112}St. Petersburg State University, St. Petersburg, Russia
\item \Idef{org113}Stefan Meyer Institut f\"{u}r Subatomare Physik (SMI), Vienna, Austria
\item \Idef{org114}SUBATECH, IMT Atlantique, Universit\'{e} de Nantes, CNRS-IN2P3, Nantes, France
\item \Idef{org115}Suranaree University of Technology, Nakhon Ratchasima, Thailand
\item \Idef{org116}Technical University of Ko\v{s}ice, Ko\v{s}ice, Slovakia
\item \Idef{org117}Technische Universit\"{a}t M\"{u}nchen, Excellence Cluster 'Universe', Munich, Germany
\item \Idef{org118}The Henryk Niewodniczanski Institute of Nuclear Physics, Polish Academy of Sciences, Cracow, Poland
\item \Idef{org119}The University of Texas at Austin, Austin, Texas, United States
\item \Idef{org120}Universidad Aut\'{o}noma de Sinaloa, Culiac\'{a}n, Mexico
\item \Idef{org121}Universidade de S\~{a}o Paulo (USP), S\~{a}o Paulo, Brazil
\item \Idef{org122}Universidade Estadual de Campinas (UNICAMP), Campinas, Brazil
\item \Idef{org123}Universidade Federal do ABC, Santo Andre, Brazil
\item \Idef{org124}University of Cape Town, Cape Town, South Africa
\item \Idef{org125}University of Houston, Houston, Texas, United States
\item \Idef{org126}University of Jyv\"{a}skyl\"{a}, Jyv\"{a}skyl\"{a}, Finland
\item \Idef{org127}University of Liverpool, Liverpool, United Kingdom
\item \Idef{org128}University of Science and Techonology of China, Hefei, China
\item \Idef{org129}University of South-Eastern Norway, Tonsberg, Norway
\item \Idef{org130}University of Tennessee, Knoxville, Tennessee, United States
\item \Idef{org131}University of the Witwatersrand, Johannesburg, South Africa
\item \Idef{org132}University of Tokyo, Tokyo, Japan
\item \Idef{org133}University of Tsukuba, Tsukuba, Japan
\item \Idef{org134}Universit\'{e} Clermont Auvergne, CNRS/IN2P3, LPC, Clermont-Ferrand, France
\item \Idef{org135}Universit\'{e} de Lyon, Universit\'{e} Lyon 1, CNRS/IN2P3, IPN-Lyon, Villeurbanne, Lyon, France
\item \Idef{org136}Universit\'{e} de Strasbourg, CNRS, IPHC UMR 7178, F-67000 Strasbourg, France, Strasbourg, France
\item \Idef{org137}Universit\'{e} Paris-Saclay Centre d'Etudes de Saclay (CEA), IRFU, D\'{e}partment de Physique Nucl\'{e}aire (DPhN), Saclay, France
\item \Idef{org138}Universit\`{a} degli Studi di Foggia, Foggia, Italy
\item \Idef{org139}Universit\`{a} degli Studi di Pavia, Pavia, Italy
\item \Idef{org140}Universit\`{a} di Brescia, Brescia, Italy
\item \Idef{org141}Variable Energy Cyclotron Centre, Homi Bhabha National Institute, Kolkata, India
\item \Idef{org142}Warsaw University of Technology, Warsaw, Poland
\item \Idef{org143}Wayne State University, Detroit, Michigan, United States
\item \Idef{org144}Westf\"{a}lische Wilhelms-Universit\"{a}t M\"{u}nster, Institut f\"{u}r Kernphysik, M\"{u}nster, Germany
\item \Idef{org145}Wigner Research Centre for Physics, Hungarian Academy of Sciences, Budapest, Hungary
\item \Idef{org146}Yale University, New Haven, Connecticut, United States
\item \Idef{org147}Yonsei University, Seoul, Republic of Korea
\end{Authlist}
\endgroup
 
\end{document}